\documentclass[journal=nalefd,manuscript=article]{achemso}
\usepackage{chemformula} 
\usepackage[T1]{fontenc} 
\usepackage[version=3]{mhchem}

\usepackage{graphicx}
\usepackage{dcolumn}
\usepackage{bm}

\usepackage{bbm}
\usepackage[utf8]{inputenc}
\usepackage[T1]{fontenc}

\usepackage{etoolbox}

\usepackage{color}

\usepackage{amssymb}
\usepackage{amsmath}

\usepackage{subeqnarray}

\allowdisplaybreaks[4]
\title[]{Topology-engineered orbital Hall effect in two-dimensional ferromagnets}

\author{Zhiqi Chen}
\author{Runhan Li}
\author{Yingxi Bai}
\author{ Ning Mao}
\affiliation{School of Physics, State Key Laboratory of Crystal Materials, Shandong University, Jinan 250100, China}

\author{Mahmoud Zeer}
\affiliation{Peter Grünberg Institute, Forschungszentrum Jülich, 52425 Jülich, Germany}
\alsoaffiliation{Department of Physics, RWTH Aachen University, 52056 Aachen, Germany}

\author{Dongwook Go}
\affiliation{Institute of Physics, Johannes Gutenberg University Mainz, 55099 Mainz, Germany}

\author{Ying Dai}
\email{daiy60@sdu.edu.cn}
\author{Baibiao Huang}
\affiliation{School of Physics, State Key Laboratory of Crystal Materials, Shandong University, Jinan 250100, China}

\author{Yuriy Mokrousov}
\affiliation{Institute of Physics, Johannes Gutenberg University Mainz, 55099 Mainz, Germany}

\author{Chengwang Niu}
\email{c.niu@sdu.edu.cn}
\affiliation{School of Physics, State Key Laboratory of Crystal Materials, Shandong University, Jinan 250100, China}
\date{\today}

\title[An \textsf{achemso} demo]{Topology-engineered orbital Hall effect in two-dimensional ferromagnets}

\keywords{orbital Hall effect, quantum anomalous Hall effect, second-order topological insulators, topological phase transation}

\begin{document}

\begin{tocentry}
	\includegraphics{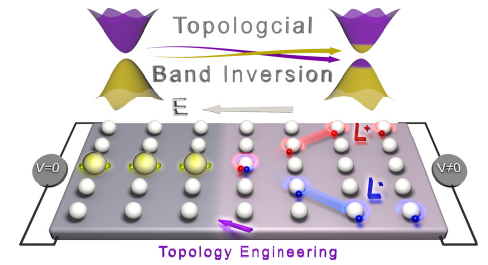}
\end{tocentry}

\begin{abstract}
Recent advances in manipulation of orbital angular momentum (OAM) within the paradigm of orbitronics present a promising avenue for the design of future electronic devices. In this context, the recently observed orbital Hall effect (OHE) occupies a special place. Here, focusing on both the second-order topological and quantum anomalous Hall insulators in two-dimensional ferromagnets, we demonstrate that topological phase transitions present an efficient and straightforward way to engineer the OHE, where the OAM distribution can be controlled by the nature of the band inversion. Using first-principles calculations, we identify Janus RuBrCl and three septuple layers of MnBi$_2$Te$_4$  as experimentally feasible examples of the proposed mechanism of OHE engineering by topology. With our work we open up new possibilities for innovative applications in topological spintronics and orbitronics.
\end{abstract}

\maketitle

The orbital angular momentum (OAM), as a fundamental degree of freedom for electrons, has recently gained renewed interest with potential applications in logic and memory devices, thereby launching the field of orbitronics~\cite{bernevig2005prl, go2017orbitronics,Go_2021}.  The field of orbitronics is currently developing very rapidly owing to its potential in resolving the problem of critical raw materials. The foundation of orbitronics is the orbital Hall effect (OHE), which refers to the emergence of electron OAM flow transverse to the external electric field~\cite{bernevig2005prl}, similar to the spin Hall effect in the realm of spintronics. Only recently, contrary to the common belief, it was shown that the magnitude of the OAM achieved out of equilibrium can be significant~\cite{sunko2017maximal,Park2011prl,sharpe2019science, tschirhart2021science}, resulting in an experimentally observable OHE in solids~\cite{choi2023observation,lyalin2023prl,sala2023prl}, present even without spin-orbit coupling (SOC)~\cite{Tanaka2008PRB,Kontani2008PRL,Kontani2009prl,Go2018prl,bernevig2005prl,Phong2019PRL,Cysne2021prl,mu2021npj}. Moreover, OHE has been revealed to serve as the fundamental origin for both spin and anomalous Hall effect~\cite{Kontani2009prl}. In experiments on magnetic layered systems, with the advances in OAM current generation, the OHE-induced orbital torques on the magnetization have been observed~\cite{ding2022orbital-torques,ding2022prl}, which may lead to new protocols for electrical control of magnetism~\cite{Dongwook2020PRR,Lee2021NC}. Naturally, once directional control of the OHE is achieved, it will become a superior alternative to the spin Hall effect, bringing revolutionary breakthroughs in various fields such as transport, information technology, and magnetic electrical control~\cite{bernevig2005prl,yuriy2023prl}. However, in spite of extensive efforts so far, an effective control of OHE still remains elusive, which presents a key challenge for practical applications of the effect.

It is generally accepted that a topological phase transition (TPT) is always accompanied by the exchange of orbital components between conduction and valence bands, referred to as band inversion~\cite{hasan2010colloquium,Qi2011,Bansi2016TBTreview}. The band inversion is a physical phenomenon that can lead to unique states at the surface $-$ as pointed out already by William Shockley back in 1939~\cite{Shockley1939PR} $-$ and to date it is widely used to conceptualize the emergence of  different topological states and topological boundary modes~\cite{Bernevig2006science,Xiao2021NRP}.  A key observation  is that a band inversion directly invokes the OAM of the inverted states, with distinct orbital components characterized by different OAM eigenvalues~\cite{Huang2022prb,Kane2018prb}, shedding light on the profound potential of band inversion, instigated by a TPT, in modulating the OHE.  While the effect of symmetry breaking on the OHE has been considered from models in non-magnetic materials~\cite{barbosa2023orbital}, magnetic materials offer much more versatility in realizing TPTs, and the control of OHE by TPTs in ferromagnets remains a great challenge with strong practical implications. 

In present work, we theoretically establish the coupling between the band inversion and OHE in 2D  ferromagnets. We emphasize that the changes of OAM resulting from band inversion significantly impact the magnitude of the orbital Hall conductivity (OHC). A multi-orbital tight-binding (TB) model is constructed to demonstrate the feasibility of attaining the proposed topology-engineered OHE with TPTs among the second-order topological insulator (SOTI), quantum anomalous Hall insulator (QAHI) and normal insulator (NI) phases in 2D honeycomb ferromagnets.  Furthermore, we identify  realistic material candidates $-$ Janus RuBrCl and three septuple layers of MnBi$_2$Te$_4$ $-$ hosting the TPTs, to validate the feasibility of topological tuning OHE. Our research offers a robust methodology to modulate the OHE, thereby setting the groundwork for novel orbitronic devices.

\begin{figure}
	\centering
	\includegraphics[width=0.9\textwidth]{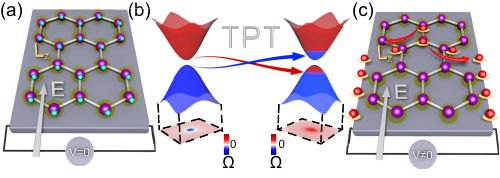}
	\caption{Sketch of the topology-engineered OHE. (a) and (c) represent the electronic orbital distribution in 2D ferromagnets for trivial and nontrivial phases without and with band inversions, respectively. The purple spheres represent atomic nuclei in a crystal, and the yellow transparent areas mark the atomic potential around the nuclei. Blue and red spheres represent electrons with zero and nonzero OAM, respectively. The yellow arrows illustrate the direction of OAM. (b) Band inversion accompanied with a TPT. The area enclosed by the dashed lines is the orbital Berry curvature in Brillouin zone.}
	\label{mechanism}
\end{figure}

In 2D ferromagnets, we are predominantly interested in the out-of-plane $z$-component of the OAM, represented by the operator $\hat{L}_z$ as illustrated in Fig.~\ref{mechanism}. Without loss of generality, we start from a trivial insulator without a band inversion, whose states around the band gap are comprised of orbitals $-$ e.g., $s$, $p_z$ and/or $d_{z^2}$ $-$ which are  the eigenvalues of  $\hat{L}_z$ with magnetic quantum number $m_z = 0$, so that the corresponding Bloch states will not carry any OAM. Often, when the band inversion happens, accompanied with a TPT, as shown in Fig.~\ref{mechanism}(b), the Bloch states of non-zero $m_z$ $-$ e.g. $p_x$, $p_y$, $d_{xz}$, $d_{yz}$ with $m_z= \pm1$ and $d_{xy}$, $d_{x^2-y^2}$ with  $m_z=\pm2$ $-$ enter the region around the gap. This provides necessary orbital complexity for enabling the electric-field mediated orbital hybridization among the states of different $m_z$, which is key to the emergence of the non-zero orbital Hall effect, Fig.~\ref{mechanism}(c). This implies that via educated orbital engineering of the band dynamics in a solid, taking place in response to various stimuli such as strain and chemical composition, we acquire a unique possibility to control the sign and magnitude of the OHE by a TPT.

\begin{figure}[h]
	\centering
	\includegraphics[width=0.7\textwidth]{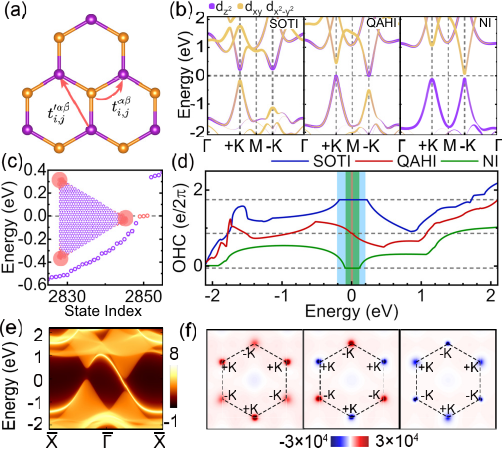}	
	\caption{ (a) Sketch of the honeycomb lattice with $C_3$ symmetry. The arrows denote the directions of hopping between relative atoms. (b) Orbitally resolved band structures for SOTI, QAHI, and NI phases of the TB model, weighted with the contribution of $d_{z^2}$ and $d_{x^2-y^2}$/$d_{xy}$. (c) Energy spectrum of a triangular nanoflake for the SOTI phase, where the occupied corner states are marked by red dots. Insets show the total charge distribution of the three occupied corner states. (d) Energy dependence of the OHC for SOTI, QAHI, and NI phases, suggesting that the OHE can be engineered via the TPTs. (e) The chiral edge state for QAHI phase with Chern number $\mathcal{C} =1$. (f) K-space distributions of orbital Berry curvatures within the insulating gap.}
	\label{model}
\end{figure}

To establish the TPT-engineering mechanism of the OHE, we employ a rich multi-orbital TB model on a 2D honeycomb lattice with the $D_{3h}$ point group symmetry ~\cite{yao2013TMD}
\begin{align}
	H =& -t_{i,j}^{\alpha \beta }\sum _{\langle i,j \rangle }\sum_{\alpha \beta }c^{\dagger }_{i\alpha }c_{j\beta }-t_{i,j}^{\prime \alpha \beta }\sum _{\langle \langle  i,j \rangle \rangle }\sum_{\alpha \beta }c^{\dagger }_{i\alpha }c_{j\beta }   \nonumber \\
	&+\sum_{i}\sum_{\alpha}(\epsilon _{i\alpha}  \mathbbm{1} + t_{m}\sigma_z )c^{\dagger }_{i\alpha }c_{i\alpha } + H_\mathrm{soc},
\end{align}
which is sketched in Fig.~\ref{model}(a). We select $d_{z^2}$, $d_{xy}$, $d_{x^2-y^2}$ orbitals and the $p_x$, $p_y$ orbitals as the basis for two sublattices, respectively. $c^{\dagger }_{i\alpha}$ and $c_{i\alpha}$ are the creation and annihilation operators for an electron with orbital $\alpha$ on site $i$.  The first and second terms represent the nearest-neighbor and next nearest-neighbor hoppings. The orbital-dependent $\epsilon _{i\alpha}$ is the on-site energy, and the magnitude of exchange field is indicated by $t_m$. The SOC interaction is introduced with $H_{soc} = t_{so}\mathbf{l} \cdot \boldsymbol{\sigma}$, where $\boldsymbol{\sigma}$ is the vector of Pauli matrices.
Figure~\ref{model}(b) presents the orbitally resolved band structures of three insulating phases under different on-site energies with $\epsilon_2 = 3.2$, $\epsilon_2 = 4.1$, and $\epsilon_2 = 4.9$~\,eV, respectively. Clearly, the conduction band minimum (CBM) and valence band maximum (VBM) are dominated by the $d_{z^2}$ and  $d_{xy}$/$d_{x^2-y^2}$ orbitals, and the orbital contributions can be effectively modified. Band inversions occur in the $+K$ and $-K$ valleys sequentially, revealing TPTs with increasing $\epsilon_2$. For $\epsilon_2 = 3.2$~\,eV, a floating edge state emerges in the insulating gap~\cite{supplemental}, which is usually considered as a key signature of the SOTI~\cite{li2022soti,bai2023engineering,mao2022orbital}. To identify its band topology, a triangular nanoflake with $C_3$ rotational symmetry is constructed and the energy spectrum is presented in Fig.~\ref{model}(c). Obviously, around the Fermi level, one observes three states (red dots) with their spatial distribution well localized at the three corners of the nanoflake, verifying the SOTI phase. 
After the band inversion at $+K$, the Chern number $\mathcal{C}$ of all occupied states indeed acquires an integer value of $\mathcal{C} = 1$, resulting in a TPT from the SOTI to QAHI, as further confirmed from the emergence of one chiral edge state in the nanoribbon displayed in Fig.~\ref{model}(e). Similarly, another TPT takes place accompanied by the band inversion at $-K$ when $\epsilon_2$ is further increased,  where the system changes from the QAHI to a NI with both the isolated corner states and chiral edge states disappearing.

According to the linear response theory, the intrinsic transverse OAM current and OHC are governed by the Fermi sea term of the Kubo formula according to
\begin{eqnarray}
	\sigma_{ij}^{z} = - e\left. \int{}_{BZ}\frac{d^{2}\mathbf{k}}{(2\pi)^{2}} \right.{\sum\limits_{n}{f_{n}\left( \mathbf{k} \right)\Omega_{n,ij}^{o,z}\left( \mathbf{k} \right)}},
\end{eqnarray}
where $\Omega_{n,ij}^{O,z}\left( \mathbf{k} \right) $ is the orbital Berry curvature, given by
\begin{eqnarray}
	\Omega_{n,ij}^{O,z}\left( \mathbf{k} \right)=
	2\hbar \mathrm{Im} \sum_{m \neq n} \frac{\left\langle u_{\mathbf{k}}^{n} | \mathcal{J}_{o,i}^{z}| u_{\mathbf{k}}^{m} \right\rangle\left\langle u_{\mathbf{k}}^{m} | \hat{\mathcal{\nu}}_{j}| u_{\mathbf{k}}^{n} \right\rangle}{(\varepsilon_{\mathbf{k}}^n-\varepsilon_{\mathbf{k}}^m)^2},
\end{eqnarray}
with $u_{\mathbf{k}}^{n}$ representing the periodic part of the Bloch state associated with the energy $\varepsilon_{\mathbf{k}}^n$, $f_{n}\left( \mathbf{k} \right)$ is the equilibrium Fermi distribution function and the velocity operator is denoted as $\hat{\nu}_j= \hbar^{-1} \partial \mathcal{H}_{k} / \partial k_j $, where  $\mathcal{H}_k$ is the Hamiltonian in momentum space. The OAM current operator is expressed as  $\mathcal{J}_{o,i}^{z}=\left\{\hat{\nu}_{i},\hat{{L}}_{z}\right\}/2$~\cite{Kontani2009prl, Bonbien2020prb}. 

In Fig.~\ref{model}(d), we plot the energy dependence of OHCs for the SOTI, QAHI, and NI phases of the model. Indeed, the OHCs are very susceptible to the topological properties, undergoing a step-wise reduction following the  TPTs. The OHC for SOTI phase displays a large magnitude of 1.74\,$e/2\pi$ in the insulating region, and stays constant throughout the band gap, i.e., SOTI phase is accompanied by a finite OHC~\cite{soti-ohc2023prl}. For QAHI phase, the value of OHC decreases to half of that in the SOTI phase, accounting to 0.87\,$e/2\pi$, and down to roughly zero for the NI phase. This is due to the fact that the $\hat{L}_z$-character of the states around the CBM and VBM is switched, accompanied by the TPTs. For example, when the TPT from SOTI to QAHI occurs, the $d_{xy}$/$d_{x^2-y^2}$ orbitals at valley $+K$ are replaced by the $d_{z^2}$ orbitals, resulting in a reversal of $\hat{L}_z$ for the CBM and VBM at $+K$. This reversal leads to a reversal in the sign of $\Omega_{n,ij}^{O,z}\left( \mathbf{k} \right)$ around $+K$ as illustrated in Fig.~\ref{model}(f). At the same time the orbital dominance of $d_{xy}$/$d_{x^2-y^2}$ states is preserved at $-K$ point which preserves the sign of the orbital Berry curvature there. This reduces the overall OHC in QAHI state by approximately half following the TPT. The situation is exactly reversed upon the second TPT when entering the NI state, where the orbital reversal and sign flip of the orbital Berry curvature occur at $-K$.  Subsequently, in the NI state the slightly negative contributions around $\pm K$ compete with the positive background contribution in the rest of the Brillouin zone, resulting in vanishing OHC.

\begin{figure*}[t]
	\centering
	\includegraphics[width=\textwidth]{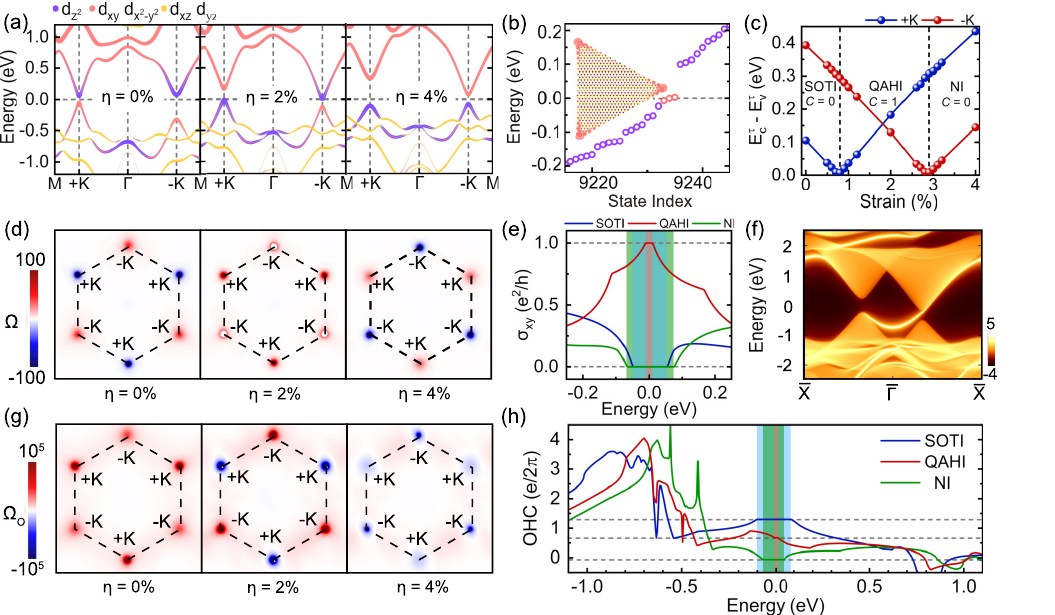}
	\caption{(a) Orbitally resolved band structures for Janus RuBrCl in (left) SOTI, (middle) QAHI, and (right) NI phases, weighted with the contribution of Ru-$d_{z^2}$, Ru-$d_{xy}/d_{x^2-y^2}$, and Ru-$d_{xz}/d_{yz}$ orbitals. (b) Energy spectrum of a triangular armchair nanoflake for Janus RuBrCl in SOTI state, where the occupied corner states are marked by red dots. Insets show the total charge distribution of the three occupied corner states. (c) Band gaps at valleys $+K$ and $-K$ as a function of tensile strain. The $k$-space distribution of (d) Berry curvatures and (g) orbital Berry curvatures for SOTI, QAHI, and  NI phases, and the corresponding energy dependence of (e) anomalous Hall conductivities  and (h) OHCs. (f) The chiral edge state of a semi-infinite RuBrCl nanoribbon for QAHI phase with Chern number $\mathcal{C} =+1$. } 
	\label{transition}
\end{figure*}

To provide a realistic manifestation of the model considerations above, we use first principles methods to identify Janus RuBrCl as the feasible candidate.	With the $C_{3V}$ point group, Janus RuBrCl exhibits a trigonal prismatic crystal field, and the five $4d$ orbitals of Ru atom split into three groups: $a_1(d_{z^2} )$, $e_1(d_{xy}, d_{x^2-y^2} )$, and $e_2 (d_{xz}, d_{yz})$. The valence electronic configuration of an isolated Ru atom is $4d^{7}5s^{1}$, and, for Janus RuBrCl, each Ru atom transfers two electrons to neighboring Br and Cl atoms, i.e., the Ru$^{2+}$ in RuBrCl has six valence electrons left, resulting in the high spin configuration with a magnetic moment of $4\mu_B$~\cite{supplemental}. 	Thus,  the band structure of Janus RuBrCl is strongly spin polarized with almost only spin-down bands residing around the Fermi level, which can be observed in the band structure without SOC, see Fig.S9~\cite{supplemental}. When SOC is switched on,  the insulating and spin-polarized character is maintained with a direct band gap of 106\,meV, and the valley degeneracy in Janus RuBrCl at the $\pm K$ points is lifted, as illustrated in Fig.~\ref{transition}(a). In addition, the spin polarization causes electrons with a definite OAM near the Fermi level to predominantly reside in the spin-down states.

The same as SOTI state in the model, the VBM at both $\pm K$ is dominated by the $d_{xy}$/$d_{x^2-y^2}$ orbitals, while the CBM at $\pm K$ is dominated by the $d_{z^2}$ orbitals, respectively. To uncover the SOTI phase of Janus RuBrCl, we focus on the $C_3$ symmetry and calculate both the topological indices $\chi^{(3)}$ and fractional corner charge $Q_c^{(3)}$, defined as~\cite{schindler2019corner,Benalcazar2019PRB,li2020corner}
\begin{eqnarray}
	\chi_{(3)}&=&\left( \left[  K_1^{(3)},K_2^{(3)}\right]  \right), \\
	Q_c^{(3)}&=&\frac{2}{3}e \left[K_1^{(3)}+K_2^{(3)}\right] mod \ 2e,
\end{eqnarray} 
where $\left[K_{1,2}^{(3)}\right]$ are given by the difference between the number of occupied bands with symmetry eigenvalue $e^{2\pi i(p-1)/3}e^{\pi i/3}$ (for $p=1,2,3$) at the $K$ and $\Gamma$ points, which are represented as \#$ K_p^{(3)}$ and \#$\Gamma_p^{(3)}$, respectively, i.e. 	
\begin{eqnarray}
	\left[K_p^{(3)}\right]=\#K_p^{(3)}-\#\Gamma_p^{(3)}.
\end{eqnarray}
The calculated topological indices and fractional corner charge are $\chi^{(3)}=(-1,2)$ and $Q_c^{(3)}= \frac{2}{3}e$, respectively, providing explicit evidence of the SOTI nature in Janus RuBrCl. Moreover, Fig.~\ref{transition}(b) presents the energy spectrum of a triangular nanoflake and the distribution of wave functions for these energy states. Three continuous states around the Fermi level can be clearly observed in the bulk gap, and the spatial distribution of wave functions is well localized at the corners of the nanoflake.

To realize the TPT discussed above, we focus on strain engineering as an effective way of modulating the electronic and topological properties~\cite{niu2015tci,lin2021strain,parker2021strain}. The magnitude of strain is described by  $\eta=(a-a_0)/a_0 \times 100\%$, where $a$ and $a_0$ represent the lattice parameters of Janus RuBrCl with and without strain. With the tensile strain increasing, as illustrated in Fig.~\ref{transition}(c), the band gaps at the $+K$ and $-K$ points undergo the successive closing and reopening process under $\eta=0.8 \%$ and $\eta=2.9 \%$, respectively, indicating an occurrence of two possible TPTs. Remarkably, as the band gaps close and reopen, the orbital character around the gaps, which is predominantly of the $d_{xy}$ and $d_{x^2-y^2}$ flavor,  is reversed between the valence and conduction bands indicating a band inversion. As a direct consequence, the sign of the Berry curvature at  $\pm K$ points flips successively, and the same positive sign at both the $+K$ and $-K$ points for $\eta=2 \%$ suggest a non-zero integral of the Berry curvature where an integer Chern number of $\mathcal{C} = 1$ is indeed obtained, as shown in Figs.~\ref{transition}(d) and~\ref{transition}(e). A TPT from SOTI to QAHI, accompanying the appearance of a band inversion at $+K$,  is thus obtained as further explicitly confirmed by the emergence of one chiral edge state illustrated in Fig.~\ref{transition}(e). Additionally, the Janus RuBrCl turns into a NI after a band inversion at $-K$ takes place. In addition to the OHE, by complementary calculations we reveal that the TPT and resulting orbital redistribution around the Fermi energy has a drastic impact on other non-equilibrium orbital properties, such as current-induced orbital polarization and spin-orbit torque, which is predominantly an orbital effect in thin 2D magnets~\cite{supplemental}.

Figure~\ref{transition}(g) presents the orbital Berry curvatures $\Omega_{n,ij}^{O,z}\left( \mathbf{k} \right)$ of the SOTI, QAHI and NI phases.  Similar to our TB model, when the TPT from SOTI to QAHI phase occurs, the reversal in the orbital character at $+K$ results in the sign reversal of $\Omega_{n,ij}^{O,z}\left( \mathbf{k} \right)$ there. As shown in Fig.~\ref{transition}(h), the OHC correspondingly decreases to the value of 0.67\,$e/2\pi$ for QAHI as compared to that of 1.30\,$e/2\pi$ for SOTI, owing to the dominance of  $m_z = \pm 2$ states at the VBM at both the $+K$ and $-K$ points for SOTI state, in contrast to the case of QAHI where this occurs only at $-K$. When the band inversion occurs at $-K$ as well, another TPT is taking place as the system changes its character from the QAHI to a NI. This drives the corresponding orbital reversal at $-K$ and overall prevalence of the $m_z=0$ character of VBM at $\pm K$, which suppresses the overall OHC to zero.

\begin{figure}[h]
	\centering
	\includegraphics[width=0.7\textwidth]{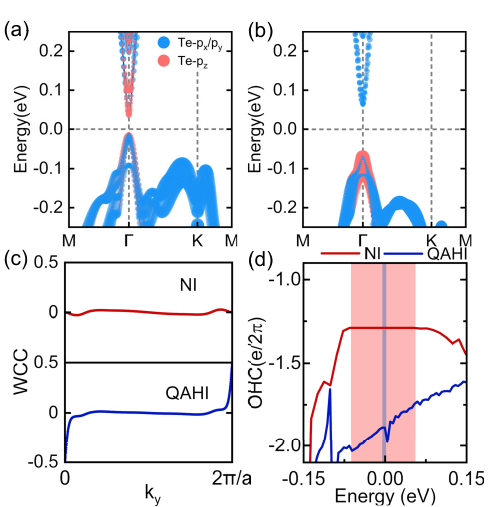}
	\caption{Orbitally resolved band structures of three septuple layers of MnBi$_2$Te$_4$ in (a) QAHI and (b) NI phases, weighted with the contribution of Te-$p_x/p_y$ and Te-$p_z$ orbitals. (c) Evolution of the WCCs and (d) energy dependence of OHC for QAHI and NI phases.}
	\label{MBT}
\end{figure}

Aiming at revealing the universality of pronounced topology-engineered OHE in 2D magnets, we turn to the case of MnBi$_2$Te$_4$, which is a prominent platform for realizing magnetic topological states, e.g., the antiferromagnetic topological insulator and QAHI have been experimentally verified in 3D and 2D MnBi$_2$Te$_4$, respectively~\cite{Gong2019CPL,otrokov2019antiferromagneticMBT,MBT2020science,gao2023MBT}. Figure~\ref{MBT}(a) displays the band structure of three septuple layers of MnBi$_2$Te$_4$ computed with SOC. The system resides in a QAHI state with $\mathcal{C} = 1$, as confirmed by the Wannier charge centers (WCCs) calculations shown in Fig~\ref{MBT}(c), consistent with the previous results~\cite{MBT2019prl}. Under compressive strain, a band gap closing and reopening occurs at the $\Gamma$ point, accompanied by the Chern number turning to zero, indicating that a TPT from QAHI to NI is achieved.

Figures~\ref{MBT}(a) and (b) illuminate the difference in orbital composition of the electronic states before and after the TPT. The VBM at the $\Gamma$ point is primarily derived from the Te-$p_x/p_y$ orbitals for three septuple layers of MnBi$_2$Te$_4$ with QAHI phase, but, in the NI phase, the Te-$p_x/p_y$ orbitals contribute to the CBM. A $p-p$ band inversion takes place here, in contrast to the $d-d$ inversion discussed above, mediated by band dynamics with $m_z= \pm 1$ character for MnBi$_2$Te$_4$. Remarkably, the band inversion results in the decreasing of $\mathcal{J}_{o,i}^{z} $ at the $\Gamma$ point, suppressing the OHC of MnBi$_2$Te$_4$ from -1.88~\,$e/2\pi$ to -1.28~\,$e/2\pi$, as illustrated in Fig.~\ref{MBT}(d). This not only marks MnBi$_2$Te$_4$ as a promising candidate for realization of topology-engineered OHE in an experimentally feasible 2D ferromagnet, but also demonstrates the possibility of manipulating OHE by topology in orbitally $p$-based systems in addition to $d$-materials.

In summary, we proposed a mechanism to effectively control the OHE via TPTs among topologically distinct phases which may even exhibit a different nature of bulk-boundary correspondence. A symmetry-constrained TB model is constructed for 2D honeycomb ferromagnets that can simultaneously host the exotic SOTI, QAHI, and NI phases, and the switchable OHE is shown to emerge as a result of the TPTs. Moreover, we demonstrated that the Janus RuBrCl and three septuple layers of MnBi$_2$Te$_4$ can realize the proposed topology-engineered OHE with $m_z = \pm 2$ and $m_z = \pm 1$, respectively, suggesting that our proposed strategy can be applied to other topological magnetic materials with non-zero OAM. Our preliminary analysis points to the fact that not only OHE is affected by the TPTs, but other orbital properties of 2D magnets are strongly influenced by the topology of the underlying electronic states as well.  We thus provide a robust methodology to modulate the OHE and generally out-of-equilibrium orbital properties of 2D ferromagnets, thereby setting the groundwork for pioneering electronic devices that utilize the combination of OHE with nontrivial topology. Owing to the origins of the OHE in crystal field properties and independence of its key features on relativistic effects, OHE can thus realize a promise of applications based on abundant light materials in a controlled way, making this phenomenon a superior alternative to spin Hall effect~\cite{bernevig2005prl,yuriy2023prl}.

\section{Method}
The calculations of the hopping terms in our Tight-binding model are carried out by MagneticTB package~\cite{zhang2022magnetictb}. The first-principles calculations are carried out by means of density functional theory (DFT) using the projector augmented wave method in the framework of generalized gradient approximation (GGA) within the Perdew-Burke-Ernzerhof (PBE)~\cite{Perdew1996GGA} functionals using the Vienna ab initio simulation package (VASP)~\cite{KressePRB1996} and full-potential linearized augmented-plane-wave method using the FLEUR code~\cite{fleur}. The cutoff energy is fixed to 500\,eV for the plane-wave basis set, and a vacuum layer of 20\,\AA\, is used to avoid interactions between nearest slabs. The strong correlation effects for the Ru-4d electrons are handled via the GGA+U method~\cite{Anisimov1991prb} within the Dudarev scheme~\cite{Dudarev1998prb}, with a value of U = 2\,eV for Ru-4$d$ electrons~\cite{Kim2016prb,huang2021prb,tian2016nano,li2022soti} and U = 5.34\,eV for Mn-3$d$ electrons~\cite{eremeev2018new,hirahara2017large}, which has credibly been used for other Ru-based monolayer structures and MnBi$_2$Te$_4$ . All the atoms are completely relaxed until the atomic forces on each atom are smaller than 0.01\,eV/\AA, and the criterion for energy convergence is set to 10$^{-6}$\,eV. The $\Gamma$-centered Monkhorst-Pack grids of $11 \times 11 \times 1$ are employed to perform the first Brillouin zone (BZ) integral. The DFT-D3 approach~\cite{grimme2010consistent,grimme2011effect} is adopted to describe the van der Waals interactions. Using the DFT perturbation theory, the phonon calculations are carried out by using the PHONOPY package~\cite{togo2015first}. Maximally localized Wannier functions (MLWFs), combining the results of first-principles calculations of FLEUR package are constructed by using the WANNIER90 code in the basis of Ru-$d$ and Cl/Br-$p$ orbitals~\cite{freimuth2008maximally}. The Monte Carlo (MC) simulations are performed via mcsolver package~\cite{liu2019magnetic}.

\begin{suppinfo}
	
	The  Supporting Information is available free of charge at 
	\begin{itemize}
		\item Supplemental Material.pdf : Details of TB model, DFT calculation, and more analyzations of Janus RuBrCl and MnBi$_2$Te$_4$, as well as other materials.
	\end{itemize}
	
\end{suppinfo}

\medskip
\begin{acknowledgement}
 This work was supported by the National Natural Science Foundation of China (Grants No. 12174220 and No. 12074217), Taishan Scholar Program of Shandong Province (Grant No. tsqn202312010), Shandong Provincial Science Foundation for Excellent Young Scholars (Grant No. ZR2023YQ001), and Qilu Young Scholar Program of Shandong University. This work was also supported by the Deutsche Forschungsgemeinschaft (DFG, German Research Foundation) $-$ TRR 288 – 422213477 (project B06), and the Sino-German research project DISTOMAT (MO 1731/10-1). We  also gratefully acknowledge the J\"ulich Supercomputing Centre and RWTH Aachen University for providing computational resources under projects  jiff40 and jara0062.
\end{acknowledgement}

%\bibliography{reference}

\end{document}